# Contribution to the local cosmic–ray flux from the Geminga supernova

P.A. Johnson

Department of Physics and Mathematical Physics, University of Adelaide,

Adelaide, SA 5005, Australia

The contribution to the local cosmic–ray flux from the Geminga supernova is calculated assuming shock acceleration to $10^{14}$eV in a remnant which was formed several $10^5$ years ago along with the Geminga pulsar. The particles are propagated to Earth using a simple diffusion model. In the region below the knee in the spectrum, it is found the supernova may contribute 10% of the local cosmic–ray flux, assuming plausible explosion parameters. The contribution to the amplitude of the anisotropy is not in conflict with the data in this energy region.

## 1  Introduction

The bright $\gamma$-ray source Geminga has puzzled astronomers for twenty years, because of its lack of emission outside the $\gamma$-ray regime. Recently, the nature of this object has finally been established with the discovery of a 237 ms pulsar in soft X-ray data [1], which was soon confirmed in EGRET [2] COS-B [3] and SAS II [4] $\gamma$-ray data at higher energies. The measurement of pulsations between 1975 and 1992 yields a steady spin-down consistent with that of an isolated pulsar, with a characteristic age of $\tau = 3.7 \times 10^5$ years [3]. From the observed $\gamma$-ray flux, the absolute upper limit on the distance is put at $d < 380$pc, whilst a Vela-like $\gamma$-ray efficiency reduces the estimate to more like $d \simeq 40$pc. Recent evidence suggests that the pulsar may be located at the nearer rather than the further end of this scale. Optical observations of its proper motion suggest a distance of 100 pc [5], and applying the polar-cap model [6] to Geminga, a distance of $d < 50$pc is implied [7]. Presumably, parallax measurements will ultimately answer the question: if the pulsar is as close as 50pc, parallax motion should be detectable.

This suggests that Geminga is the second closest pulsar known (the other being the recently discovered PSR J0437-4715, a much older object [8]), and its existence implies that a supernova exploded nearby a few $10^5$ years ago. A supernova of typical energy with a remnant of this age accounts very well for the current soft X-ray background [9], and it has recently been suggested that this nearby cataclysm may be responsible for the local bubble in the Interstellar Medium [10]. The local bubble is also coincident with a roughly ellipsoidal shell of turbulent plasma, which has been interpreted as the shock wave from a supernova. The centre of this shell lies about 45pc distant in the approximate direction of the pulsar [11], and the sun lies in its outer fringes. Typical pulsar space velocities suggest that Geminga may have travelled 50-100pc from the site of the explosion [10], and since the inferred distance travelled normal to the galactic plane is only 15-30pc, there is scope for the pulsar having travelled a considerable distance towards or away from us. It is certainly conceivable that the explosion occured at a distance of 45pc. Fortunately, this calculation is fairly insensitive to the distance of the supernova, and here we will adopt a value of 100pc where necessary.

Young supernova remnants are thought to be capable of efficiently accelerating cosmic–rays up to $10^5$ GeV[12], and possibly $10^6$ GeV [13], and if this is the case, the contribution to the ambient cosmic–ray intensity from the Geminga remnant must be very important given its proximity to Earth. Here, we assume acceleration and subsequent diffusion from the remnant, and find that the cosmic–rays may constitute a small fraction of the observed local cosmic–ray flux just below the "knee" in the spectrum.

## 2 Cosmic–Ray Acceleration and the Flux at Earth

Neutron stars are formed in Type II supernova, where several solar masses of material are ejected at $\simeq 10,000$ km s$^{-1}$ releasing more than $10^{51}$ erg of kinetic energy [14]. Cosmic–Rays are efficiently accelerated in the shock formed in the blast phase of the supernova, which lasts for about $10^3$ yr, until the ejecta has swept up an amount of material equal to its own mass, at a radius of 10 pc or so [12]. Weaker acceleration persists into the adiabatic Sedov phase which ends when the ejecta velocity becomes subsonic and the shock decays, after about $10^6$ yr, during which the shock will have travelled typically 100 pc. Numerical simulations show that whilst the highest energy particles may achieve several $10^5$ GeV in the blast phase, less than a further factor of two increase in energy is achieved in the subsequent Sedov phase [12][15][13]. We assume here the maximum energies are essentially reached in the first phase.

At the end of the blast phase, the shock radius will still be small compared with the distance to the site of the original Geminga supernova, and since this phase is brief compared with the current age of the remnant, we assume point source production in space and time. The particles are subsequently propagated to Earth using simple diffusion. The particles will in fact be contained to some degree near to the shock until they are released at later times, but the simple assumption employed here serves as a lower limit to the estimate of the

cosmic–ray flux. The actual flux will be somewhat higher, especially as the decaying shock is now in the region of the Earth.

The diffusion coefficient of a proton of energy $E_9$ GeV in the galaxy is estimated [16] as

$$D = 2 \times 10^{28} E_9^{0.3} \quad \text{cm}^2\text{s}^{-1}$$

from observations of secondary nuclei, although the exponent on $E_9$ in this equation is not well known, and values in the range 0.3–0.6 are possible [17]. Assuming the lower value, the mean distance travelled by a proton since acceleration to the present time is then

$$<R> = \sqrt{2D\tau} = 230 E_9^{0.15} \quad \text{pc}.$$

Thus, unless the remotest estimates of the distance to the supernova are correct, we see that even 1 GeV energy protons have had time to diffuse to Earth.

Assuming point-source diffusion, the probability density of finding a particle at a distance $d$ from the supernova is given by (e.g. [17])

$$G = \frac{1}{8(\pi D\tau)^{3/2}} \exp\left(\frac{-d^2}{4D\tau}\right)$$

which reduces to

$$G \simeq 2 \times 10^{-64} E_9^{-0.45} \quad \text{cm}^{-3}$$

since $d \gg \sqrt{2D\tau}$ in the energy range of interest. The production spectrum of cosmic–rays at the shock front is a power law [18] $\frac{dN}{dE} \propto E^{-2}$ containing an equal amount of energy per decade, from which we get the intensity at Earth,

$$I = G \frac{c}{4\pi} \frac{dN}{dE} = 0.05 \xi E_{51} E_9^{-2.45} \quad \text{cm}^{-2}\text{s}^{-1}\text{sr}^{-1}\text{GeV}^{-1}$$

where the kinetic energy of the ejecta is $10^{51} E_{51}$ ergs, of which we assume a fraction $\xi$ emerges in cosmic–rays over the interval $1 \to 10^5$ GeV. This simple assumption is supported by numerical work [15] which finds that, taking into account adiabatic losses, up to 20% of the ejecta kinetic energy is given to cosmic–ray protons in a spectrum described by a power law with an index of –2.01.

The intensity is shown superimposed on a recent review of the observed cosmic–ray spectrum given in Fig.1. The spectrum is shown for a typical supernova with $E_{51} = 1$, and with an efficiency $\xi = 0.2$ which is in the middle of a range quoted by other work (0.1,0.5)[13]. The shading represents the effect of varying the energy exponent of the diffusion coefficient from 0.3 to 0.6 It is also extended to indicate the approximate effect of the maximum energy being as high as $10^6$eV.

It is clear that the using this simple model, the cosmic–ray flux from the Geminga supernova can at best account for 10% – and more likely only 1% – of the observed flux near to the knee, and less at lower energies.

## 3  Anisotropy

The amplitude of anisotropy expected from a point source of cosmic–rays is

$$\delta = \frac{3D}{c}\frac{|\nabla n|}{n}$$

which reduces to

$$\delta = f(E)\frac{3d}{2c\tau}$$

if we express the density of cosmic–rays from Geminga as a fraction $f(E)$ of the total observed density. Taking $\tau$ as the characteristic age of the neutron-star and $d = 100$pc, we obtain $\delta = 10^{-3} f(E)$. The observed amplitude is constant around $10^{-3}$ in the region $10^3$ to $10^6$GeV, with a phase near to a right ascension of $3^h$[19][20]. Near to the knee, where we have seen that $f(E)$ is at best about 0.1, the expected anisotropy from Geminga is therefore below that required to completely account for the observed value, and is consistent with a picture in which the observed anistropy arises due to motion along the local magnetic field. Nevertheless it is worth noting that the supernova lies roughly in the direction of the observed phase of the anisotropy. The high value of the proper motion of the neutron-star implies that it has travelled 17 degrees from the site of the original supernova, and projecting back the trajectory of the neutron star, the right ascension of the progenitor is estimated as $5^h40^m$ [5]. Whilst this is few hours different than the observed phase, since the remnant has a large apparent size, any asymmetry or clumpiness in the supernova could probably produce the required phase.

# 4  Discussion

A corollary of this model is that we would expect some temporal variation in the cosmic–ray flux arriving at Earth over the last several $10^5$ years. Measurements of $^{36}Cl$ in meteorites indicate however that the average cosmic–ray flux has remained constant to within a factor of two in the last $10^6$ years [21]. This is not necessarily in conflict with the model presented here, because such measurements reflect the low energy component of the cosmic–ray flux, whilst the model here indicates that cosmic–rays from Geminga might be most important near to the knee. There is some evidence for time variability of the cosmic–ray flux however. An enhancement is seen in the concentration of $^{10}Be$ in several ice cores and sea-sediments around the world at an agreed date of 35,000 years ago, the strength of which is not explainable by fluctuations in the Earths dipole field or the solar output [22]. A more likely hypothesis has been proposed, that a supernova exploded nearby in the last $10^5$ years or so, and that the $^{10}Be$ spike is the signature of the cosmic–ray enriched shock wave passing through the Earth [23]. As we have seen, the shock wave from Geminga will be in the region of the Earth after several $10^5$ years, and could quite conceivably have passed by us 35,000 years ago. Though the calculated cosmic–ray flux is not sufficient to account for this spike in the simple model, this coincidence could be a clue, possibly indicating strong trapping of cosmic–rays at the shock.

The conventional view of course is that cosmic–rays permeate the whole of the galaxy, but there is evidence to suggest that the cosmic–rays that we observe are of relatively local origin. This has been used [24] to suggest that cosmic–rays are trapped by a local magnetic supershell, and no evidence from composition or anisotropy measurements was found which precluded the production and trapping of cosmic–rays in such a region.

It has been shown here that if the simple diffusion picture is correct, particle acceleration in the Geminga supernova could be responsible for a small fraction of the present-day cosmic–rays just below the knee, but cannot account for the whole of the cosmic–rays in that energy region. However, if trapping of the cosmic–rays near to the shock extends into the Sedov phase, the density of particles could be significantly enhanced. Since acceleration does persist into this phase, albeit at a reduced rate, it could be the case that the magnetic field structure in the shock is sufficient to detain the cosmic–rays, which will alter the simple picture presented here.

*Acknowledgements*

I thank L. O'C. Drury, R. J. Protheroe and an anonymous referee for useful comments.

Figure 1: The cosmic–ray spectrum above $10^3$ GeV from a review by Stanev[25]. The intensity has been multiplied by $E_9^{2.75}$ to flatten the spectrum and enhance its features. The expected flux of cosmic–rays from the Geminga supernova is superimposed. The shading denotes the effect of uncertainty in the galactic diffusion coefficient.

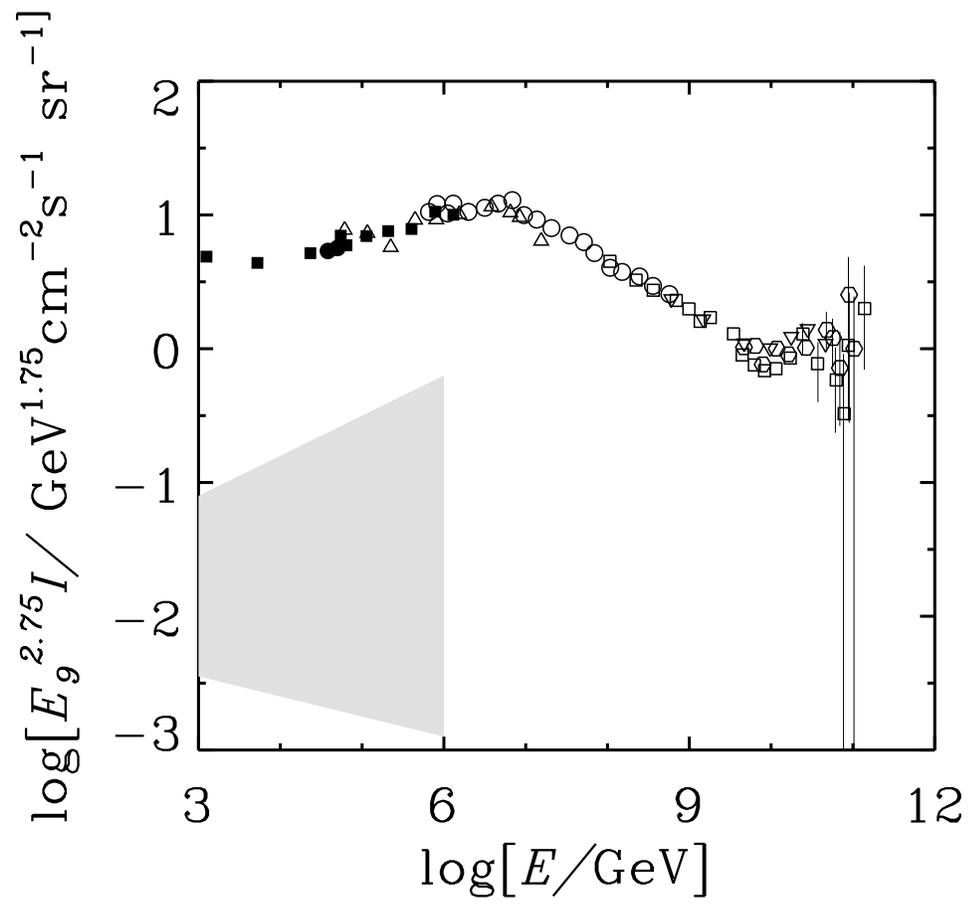

Figure 1: from P. A. Johnson 'Contribution to the local cosmic–ray flux from the Geminga Supernova'